\documentclass[paper]{JHEP3} 

\usepackage{epsfig,multicol,bbm}

\newcommand\fverb{\setbox\fverbbox=\hbox\bgroup\verb}
\newcommand\fverbdo{\egroup\medskip\noindent%
			\fbox{\unhbox\fverbbox}\ }
\newcommand\fverbit{\egroup\item[\fbox{\unhbox\fverbbox}]}
\newbox\fverbbox

\newcommand{\be}{\begin{eqnarray}}
\newcommand{\ee}{\end{eqnarray}}

\title{Large scale alignment anomalies of CMB anisotropies:
a new test for residuals applied to WMAP 5yr maps}

\author{Alessandro Gruppuso
\\
	INAF/IASF-BO, Istituto di Astrofisica Spaziale e Fisica Cosmica di Bologna,\\
	via Gobetti 101, I-40129 Bologna, Italy
\\
	INFN, Sezione di Bologna,
	Via Irnerio 46, I-40126 Bologna, Italy
\\
	E-mail: \email{gruppuso@iasfbo.inaf.it}}

\author{Carlo Burigana
\\
	INAF/IASF-BO, Istituto di Astrofisica Spaziale e Fisica Cosmica di Bologna,\\
	via Gobetti 101, I-40129 Bologna, Italy
\\
	Dipartimento di Fisica, Universit\`a degli Studi di Ferrara, \\
	via Saragat 1, I-44100 Ferrara, Italy
\\
	E-mail: \email{burigana@iasfbo.inaf.it}}

\received{\today} 		
\accepted{\today}		

\abstract{We analyze the alignment of the low multipoles (quadrupole and octupole) of various maps of the WMAP 5yr release:  the CMB maps obtained with ILC and MCMC methods, the CMB map in the V band after foreground reduction, and, for comparison, the (not cleaned) V band map.
We study how much this alignment is polluted by residuals on the Galactic region. Among the 
considered maps, the WMAP-ILC turns out to be the most clean map from the point of view of the proposed test.
This result has been found studying the redistribution (due to the masking process) of each bin of the probability distribution functions of the alignment estimators.
By construction, our method, feasible through Monte Carlo simulations, works for any possible mask adopted in the analysis of data from current and forthcoming CMB anisotropy experiments
and it can only exclude that the considered map is clean.}

\keywords{CMBR theory, CMBR experiments}

\begin{document} 


\section{Introduction}
\label{intro}

The anisotropy pattern of the cosmic microwave background
(CMB), obtained by Wilkinson Microwave Anisotropy Probe
(WMAP), probes cosmological models with unprecedented precision (see \cite{dunkley,komatsu}
and references therein). Although WMAP data are largely consistent
with the concordance $\Lambda$ cold dark matter ($\Lambda$CDM) model, there
are some interesting deviations from it, in particular on the largest
angular scales. 
They can be divided in the following categories.
{\it 1) Lack of power at large scales}.
The angular correlation function is found to be uncorreleted (i.e. consistent with $0$) for angles larger than $60$ degrees. In \cite{Copi2008, Copi2006} 
it has been shown that the probability associated to this event is low as $0.15 \%$. 
Still in this category we mention the surprisingly low amplitude of the quadrupole term of the angular power spectrum (APS), already found by Cosmic Background Explorer (COBE) \cite{Smoot1992,Hinshaw1996}, and now confirmed by
WMAP \cite{dunkley,komatsu}. 
{\it 2) Unlikely alignments of low multipoles}.
An unlikely (for a statistically
isotropic random field) alignment of the quadrupole and the octupole is
described in reference
\cite{copi2004,weeks,tegmark2003,schwarz2004,land2005,vale2005}. 
Moreover, both quadrupole and octupole align with the CMB dipole
\cite{Copi2006}. 
Other unlikely alignments are described in \cite{abramo,wiaux,vielva2007}. 
{\it 3)~Hemispherical asymmetries}.
It is found that the power coming separately from the two hemispheres (defined by the ecliptic plane) 
is too asymmetric (especially at low $\ell$) \cite{Eriksen,hansen}.
{\it 4) Cold Spot}. In Ref.~\cite{vielva} it is reported a detection of a non
Gaussian behaviour in the southern hemisphere with a wavelet
analysis technique (see also \cite{cruz}).

It is still unknown whether these anomalies come from fundamental
physics or whether they are the residual 
of some not perfectly removed astrophysical foreground
or systematic effect \cite{abramo2,Naselsky:2006mt,Inoue2006,Inoue2007,cooray2005}. 
As an example of the latter kind, in references \cite{DSCburigana,DSCgruppuso} it is presented a study about the impact
of the dipole straylight contamination
on the low amplitude of the quadrupole and on the low $\ell$ alignments
for {\it Planck}~\footnote{http://www.rssd.esa.int/planck}
characteristics and capabilities.
Many efforts have been dedicated to the development of methods aimed at the discrimination
between spurious and cosmological effects
\cite{Coles:2003dw,Park:2006dv,Naselsky:2007gt,Chiang:2007rp}.
The still open question about the origin of such anomalies has attracted a lot of interest
in the last few years. 
In \cite{Luminet:2003dx,Weeks:2006rr,Contaldi:2003zv,Moroi:2003pq,Campanelli2006,Gruppuso2007,Finelli:2005zc,Barrow1997,Gosh2007,Jaffe2006,de OliveiraCosta:2003pu} there are some papers about the possible 
generation of the low $\ell$ anomalies in the context of some specific models of fundamental physics.
In \cite{Bunn} it is claimed that no model of contamination that is statistically independent of the source of the primary CMB anisotropy, can explain this large-scale
power deficit. In other words, if a contamination (not taken into account) 
is responsible for the lack of power it must have a correlation with the primary CMB anisotropy.

\begin{figure}
\includegraphics[width=47mm, angle=90]{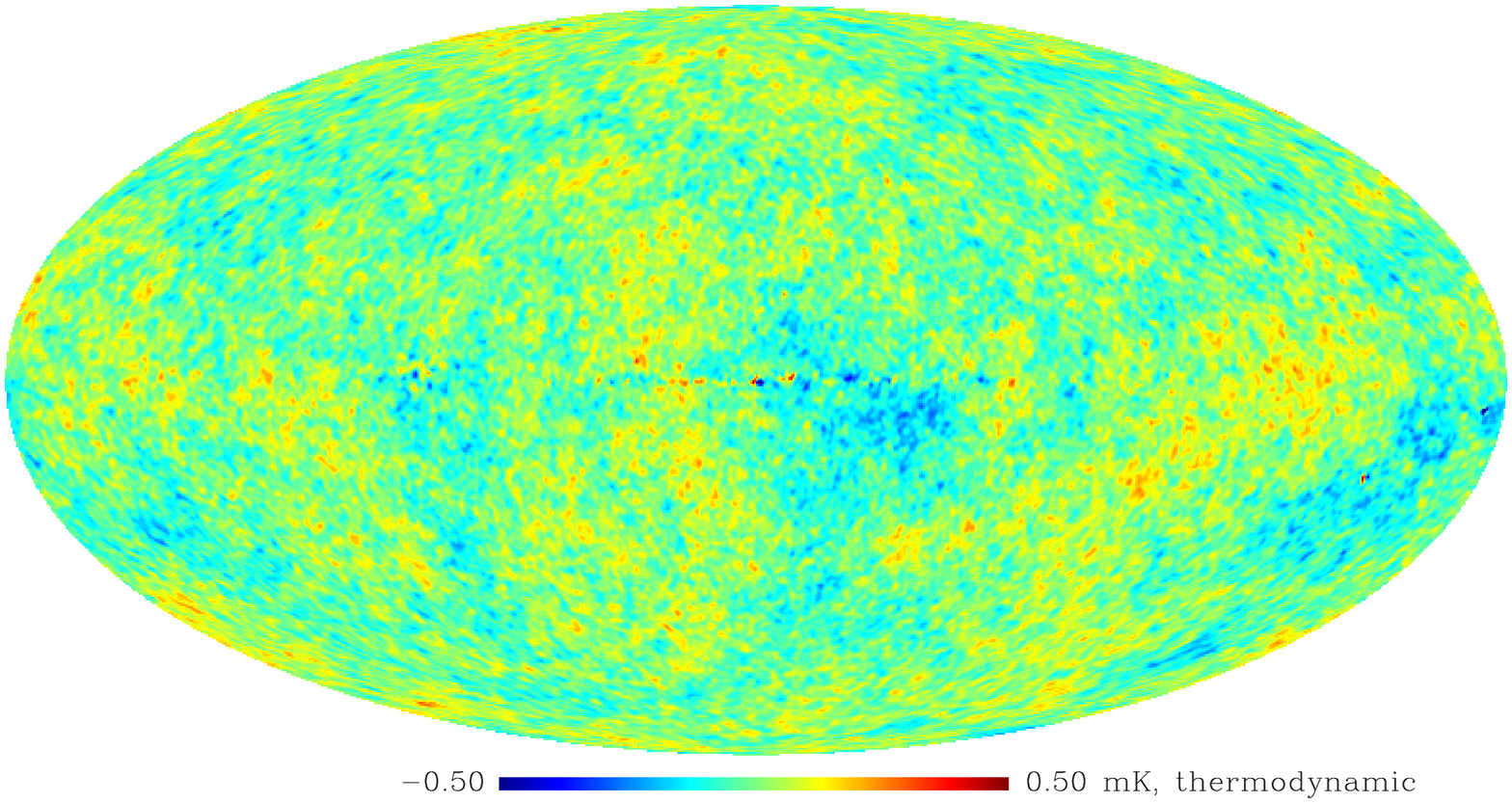}
\includegraphics[width=47mm, angle=90]{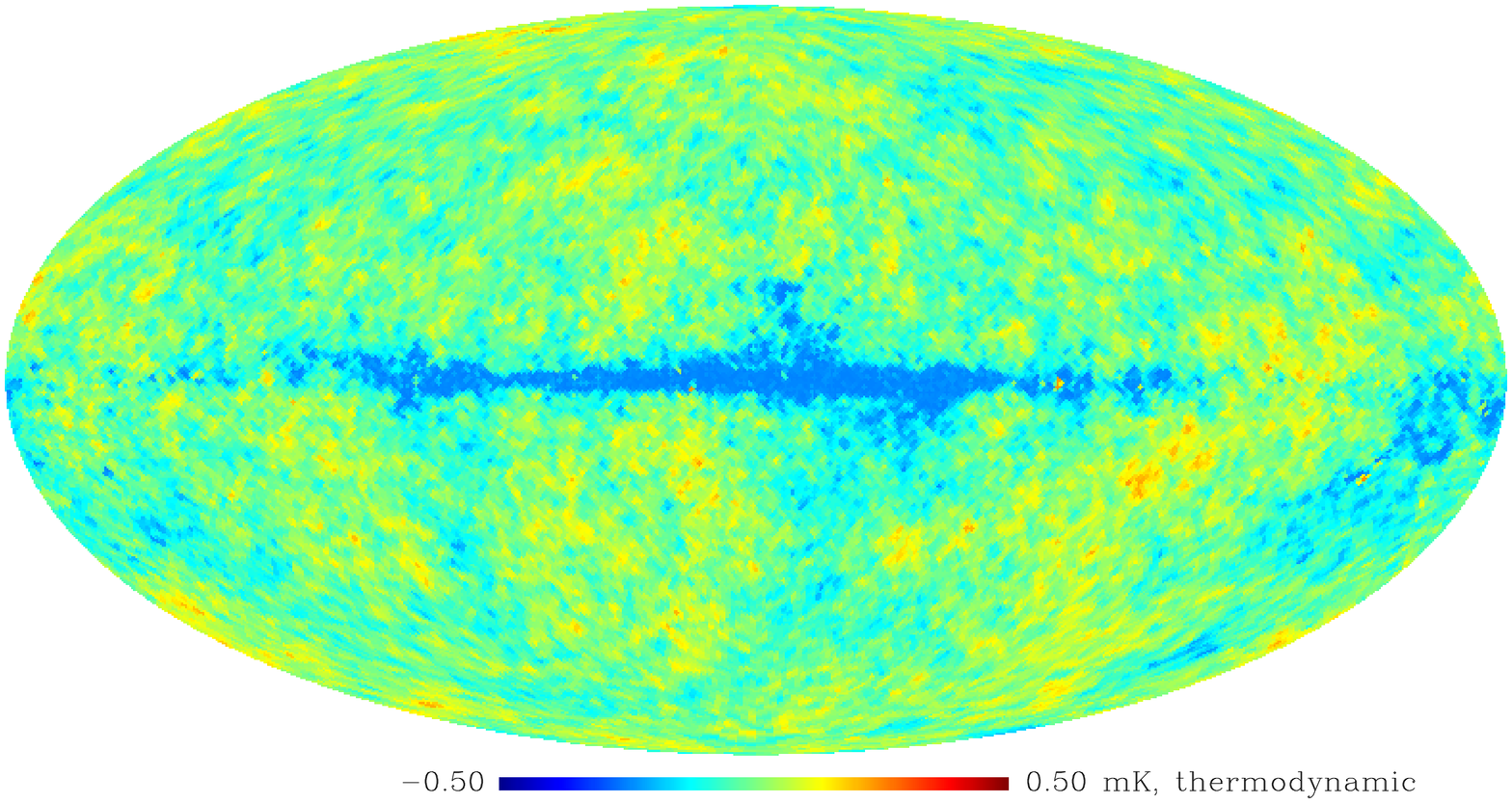}

\includegraphics[width=47mm,, angle=90]{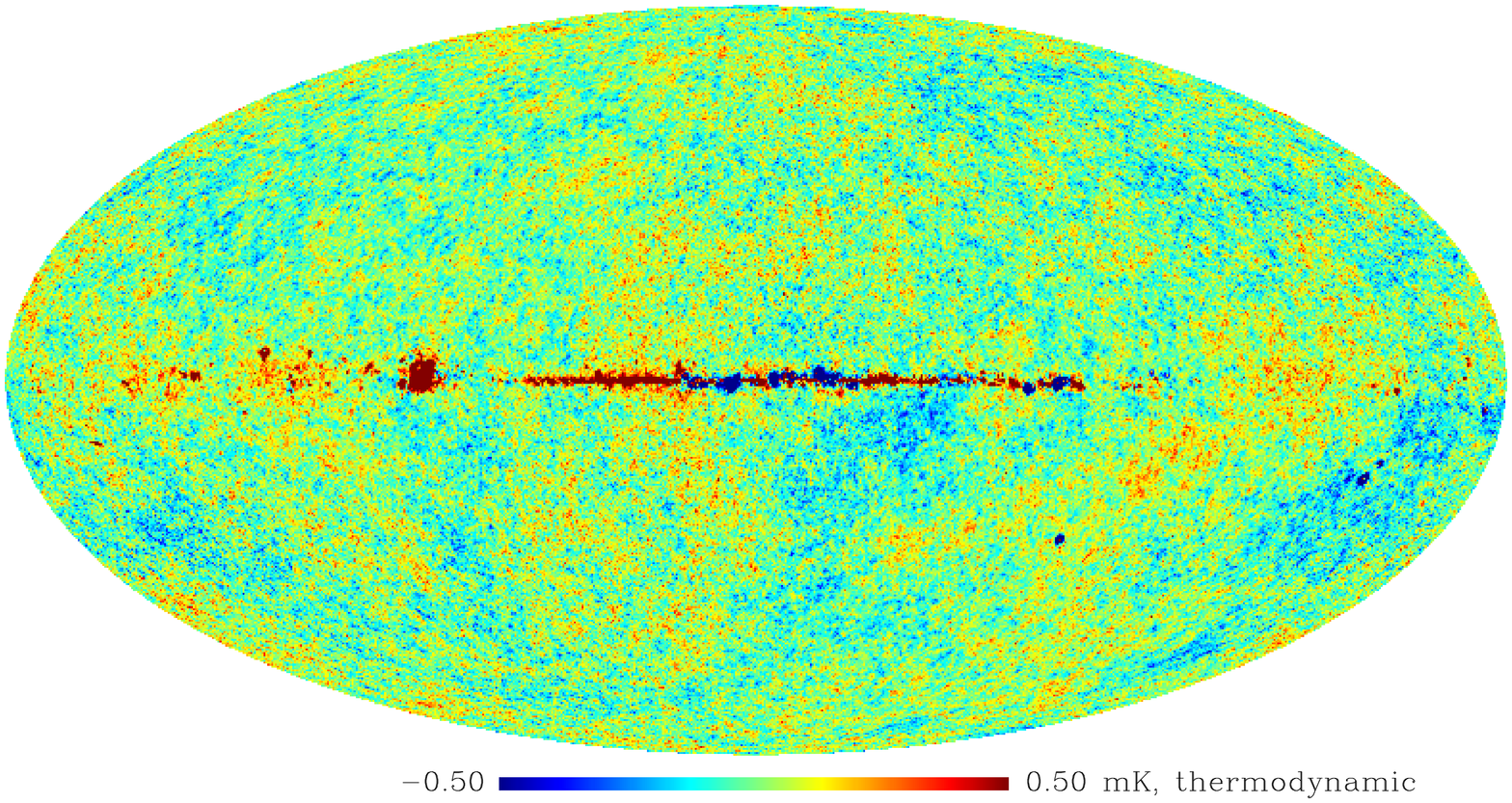}
\includegraphics[width=47mm, angle=90]{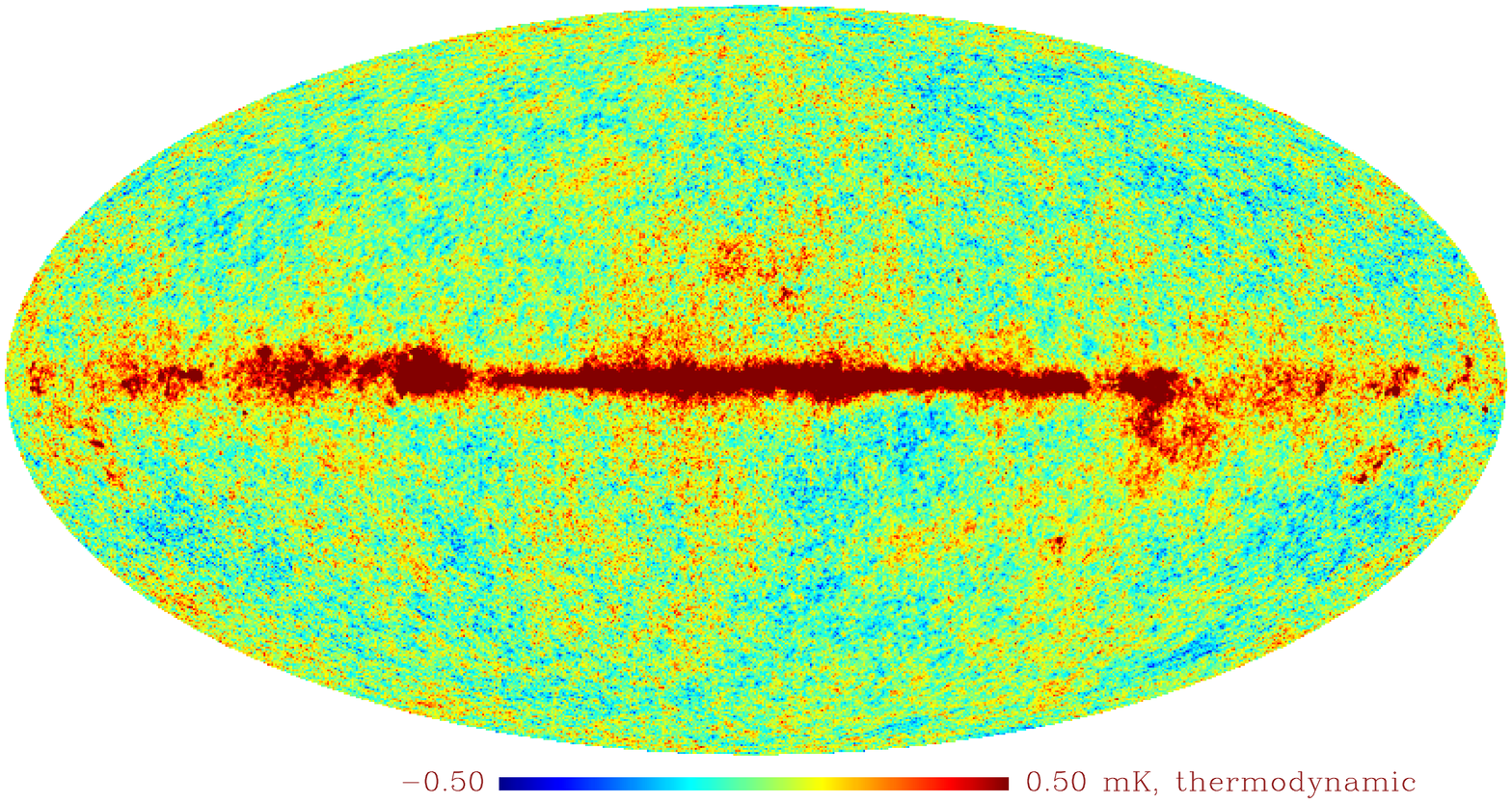}
\caption{Mollweide projections of the considered maps.
ILC map (upper left), MCMC map (upper right), foreground reduced V band map (lower left), V band map (lower right).
The number of pixels in each map is $N_{pix}=3145728$ except for the MCMC map where $N_{pix}=49152$.
A part the V band map that is taken into account as test case, all the other maps are only CMB maps obtained with different techniques.
See also the text.}
\label{ilc}
\end{figure}

In the present paper we focus on the second of the aforementioned list of anomalies, i.e. on 
the unlikely alignments of low multipoles (quadrupole and octupole). 
We take into account various maps of the WMAP 5 year release:  the ILC map, the MCMC map, the (not cleaned) V band map and the foreground reduced V band maps (see Fig.~\ref{ilc}). 
The ILC (Internal Linear Combination) map, 
available at pixel size of 6.87~arcmin,
has been obtained from a weighted 
linear combination of the five intensity maps
at the various WMAP frequencies 
smoothed to 1~degree (FWHM) resolution, 
with weights chosen to maintain the CMB 
anisotropy signal while minimizing the Galactic 
foreground contribution in different 12 regions covering the whole sky.
The WMAP team believes that it is suitable 
for analyses on angular scales greater than about 10 degrees.
The MCMC (Monte Carlo Markov Chain) map is again smoothed to 1~degree 
resolution but with a pixel size of 54.97~arcmin. From the MCMC fit
the WMAP team derived anisotropy maps for the Galactic foreground 
diffuse components and that for the CMB component, considered in 
this work. We exploit also the foreground reduced 
(i.e. the CMB anisotropy map derived with a subtraction of the 
foreground components using a Foreground Template Model) map in the V 
band, the WMAP frequency channel where the original level of 
foreground is minimum. For comparison, we consider also the map in 
the V band without any kind of foreground subtraction. These two last 
maps centred at 61~GHz have the original beam resolution of 19.8~arcmin
and are provided at a pixel size of 6.87~arcmin.
Clearly, since we will analyse the large scale properties
of the sky, the fact that the above maps have been 
provided at different resolutions and pixel sizes do not have any
significant impact for this work. 
All these products are publicly available at LAMBDA web site~\footnote{http://lambda.gsfc.nasa.gov/} where further information can be found.

In this paper we address the impact of residuals (that are unavoidably present in the considered maps) on the estimators for alignments of low multipoles.
Uncertainties can be potentially introduced in each step of data analysis needed to generate the CMB map. For example, it is known that it is difficult to perform an accurate component separation close to the Galactic plane.
Therefore we propose a new consistency test that can be capable of detecting the effects of the residuals on
the estimators for the alignments at low multipoles. 
The basic idea of this check is the comparison between the value of the estimator of the map under analysis with the 
values of the estimator of the simulated maps that are unlikely (or likely, depending on the context) at the same level of the considered map.
The application of the proposed test to the not cleaned V band map, certainly not meaningful for a cosmological analysis, 
is performed in order to verify the capability of the method to identify spurious contributions in a known dirty case.

The problem of minimizing potential residuals in a CMB map is crucial for cosmological analyses.
In \cite{bielewicz_2} it is proposed a way to limit the impact of foreground contamination
present in the all sky maps adopted for the low $\ell$ analysis of the alignments.
This new method makes use of a power equalization filter \cite{bielewicz_1}
and it has been proposed for a better control of residual foregrounds and therefore 
for a potentially more robust cosmological analysis.
However, even if this (or some other) component separation method could work perfectly, 
residuals can still be present because of uncertainties from potential
systematics and from other stages in the data analysis.
In \cite{deOliveiraCosta:2006} the $a_{\ell m}$ (coefficients of the expansion over the basis of the Spherical Harmonics) of the CMB are obtained with a minimum variance method in case of incomplete sky coverage, non uniform noise and foreground contamination.
Unfortunately, if the considered mask excludes more than about $10\%$ of the sky then the error bars of the $a_{\ell m}$ coefficients become too large.

In this paper we adopt a drastic approach and mask the Galactic region
(with masks of various size \footnote{In 2 cases out of 3, larger than $10\%$ of the sphere.}).  
Since the Multipole Vectors expansion (that is the mathematical tool that we will use to define
alignments \cite{copi2004,weeks}) is doable only over the all sky
(i.e. over the full set of the orthonormal Spherical Harmonics),
we do not look for a new basis defined in the uncut region but
we screen the information coming from the Galactic region, setting 
the value of the pixels falling into the area we do not want to consider, 
to a fiducial value (namely $0$).
In Section \ref{methodology} it is shown how to consistently perform the analysis
and how to use the masking process in order to analyze the
contamination present in the masked area
(i.e. the proposed test)
\footnote{As it will be clearer in Section \ref{methodology}, we do use the masking process as a tool to define a test for residuals and to infer conclusions about the all sky maps
but we do not use the scientific analysis on the masked maps to infer conclusion  about the all sky maps.}.
Technically, we study the redistribution (due to the masking process) of each bin of the probability distribution functions 
of the alignment estimators.
The used technique can indeed only exclude that the considered map is clean in some region but it cannot state that the map is clean in that region since a possible residual could be
in principle compatible even with random realizations. 
By construction, our method works for any possible mask suited to exclude regions that are possibly affected by various kinds of contamination.

The paper is organized as follows: in Section \ref{methodology}
we present the adopted methodology and describe the performed simulations and the proposed test,
in Section \ref{results} we show the obtained results,
and in Section \ref{conclusions} we draw our conclusions.

\section{Methodology}
\label{methodology}
\subsection{Multipole Vectors}
\label{multipolevectors}

The alignment of multipoles can be defined
using a new representation of CMB anisotropy maps
where the $a_{\ell m}$ 
are replaced by vectors \cite{copi2004,weeks}.
In particular, each multipole order $\ell$ is represented by $\ell$
unit vectors and one amplitude $A$
\be
 a_{\ell m} \leftrightarrow A^{(\ell)}, \hat u_1, \, \cdot \cdot \cdot \, , \hat u_{\ell}
\, .
\label{association}
\ee
Note that the number of independent objects is the same in the l.h.s and r.h.s. of equation (\ref{association}):
$2 \ell +1$ for $a_{\ell m}$ 
equals $3 \ell$ (numbers of components of the vectors) $+1$ (given by $A^{(\ell)}$) $-\ell$ (because there are $\ell$ constraints due to the normalization conditions of the vectors).

Equation~(\ref{association}) can be understood 
starting from this observation \cite{weeks}:
if $f$ is a solution of the Laplace equation
\be
\nabla^2 f =0 
\, ,
\ee
where $\nabla^2=\partial_x^2+\partial_y^2+\partial_z^2$ in Cartesian coordinates,
then it is possible to build a new solution $f^{\prime} $ applying a directional derivative to $f$
\be
\nabla_{\vec u} f \equiv \vec u \cdot \nabla f = f^{\prime}
\, , \; \; \; \; \; \; \nabla^2 f^{\prime} =0
\, ,
\ee
with the gradient $\nabla = (\partial_x,\partial_y,\partial_z)$.
This happens because the two operators $\nabla^2$ and $\nabla_{\vec u}$ commute. 
Maxwell \cite{maxwell} repeated this observation $\ell$ times considering the $1/r$ 
potential as starting solution.
Here $\vec r = (x,y,z)$ and 
$r =\sqrt{\vec r\cdot\ \vec r} = \sqrt{x^2 + y^2 + z^2}$.
In this way, one obtains
\begin{equation}
    f_\ell(x,y,z)
    = \nabla_{\vec u_\ell} \cdots \nabla_{\vec u_2} \nabla_{\vec u_1} \;
    \frac{1}{r} \, .
    \label{Maxwell}
\end{equation}
Observe the simple pattern
that emerges as we apply the directional derivatives one at a
time:
\begin{eqnarray}
    f_0 &=& \frac{1}{r}\nonumber\\
    f_1 &=& 
\frac{(-1)(\vec u_1 \cdot \vec r)}{r^3}\nonumber\\
    f_2 &=& 
\frac{(3\cdot 1) (\vec u_1 \cdot \vec r)(\vec u_2 \cdot \vec r) + r^2(-\vec u_1 \cdot \vec u_2)}{r^5}\nonumber\\
    f_3 &=& 
\frac{(-5\cdot 3\cdot 1) (\vec u_1 \cdot \vec r)(\vec u_2 \cdot \vec r)(\vec u_3 \cdot \vec r) + r^2(...)}{r^7} \, .\nonumber 
\end{eqnarray}
The (...) stands for a polynomial which we do not write 
explicitly, being not relevant to the current purposes.

Moreover, writing $f_{\ell}$ in spherical coordinates once $r$ is set to $1$, one finds
the following property
\be
\tilde \nabla^2 f_{\ell}(1,\theta,\phi) = \ell (\ell +1) f_{\ell}(1,\theta,\phi)
\, ,
\ee 
where $\tilde \nabla^2$ is the angular Laplace operator defined as
\be
\tilde \nabla^2 =
-\left[ {1\over \sin \theta}\, \partial_{\theta} \left( \, \sin \theta \, \partial_{\theta}\right)
+ {1\over \sin^2 \theta} \, \partial_\phi^2 \right]
\, \label{anglaplace}.
\ee 
In other words $f_{\ell}(1,\theta,\phi)$ is eigenfunction of the angular part of the Laplace operator
with eigenvalue given by $\ell (\ell +1)$. 
This is nothing but the definition of spherical harmonics $Y_{\ell,m}$ 
(e.g. see \cite{sakurai}).
Therefore, for every $\ell$ we can write
\be
A^{(\ell)} f_{\ell}(1,\theta,\phi) = \sum_{m=-\ell}^{\ell} a_{\ell m} Y_{\ell m}(\theta, \phi)
\, ,
\label{identification}
\ee
where the amplitude $A^{(\ell)}$ has been inserted because of normalization purposes.
Equation~(\ref{identification}) makes evident the association represented by 
equation~(\ref{association}).
From equation~(\ref{identification}) it is possible to write down the set of equations
that has to be solved to pass from $a_{\ell m}$ to multipole vectors.
In order to see that this set is solvable we count the equations 
and the unknowns involved in this set.
From equation~(\ref{identification}) we have $2 \ell +1$ equations (one 
equation for each independent $a_{\ell m}$
\footnote{In fact we would have $4 \ell  +1$ equation because each $\ell$ different from $0$ has a real and imaginary part. But considering that $a_{\ell m}$ with $m>0$ are related to those with $m<0$) 
through $a_{\ell m}^{\star} = (-1)^m a_{\ell -m}$ we are left with $2 \ell +1$ equations (i.e. the expression must be real).}) plus $\ell$
equations from the normality conditions of the vectors 
(i.e. $\vec u_{i} \cdot \vec u_{i} =1 $ where $i$ runs from $1$ to $\ell$).
Therefore the total number of independent equations is $3 \ell +1$. 
This is also the number of unknowns because we have $3$ unknowns for each vector plus
$1$ given by the amplitude $A^{(\ell)}$. This shows that the set is solvable.

One of the advantage of Multipole Vectors representation is that from 
these unit vectors one can easily construct scalar quantities that 
are invariant under rotation. 
Note that is not equally easy to obtain scalar quantities directly 
from the $a_{\ell m}$ coefficients since they depend on the 
coordinate system. 
For a more detailed explanation of equation (\ref{association})
and of the properties of that association
see for example references \cite{copi2004,weeks, abramo, DSCgruppuso}.

Unfortunately, an explicit analytical expression for the association
given in equation (\ref{association}) is possible only for $\ell =1$.
For $\ell \neq 1$ numerical methods are needed~\footnote{Indeed, for 
$\ell=2$ it is possible to obtain the multipole vectors
computing the eigenvectors of a symmetric and traceless tensor representing the quadrupole, see \cite{landmagueijo,dennis}.}.
%
The Copi et al.'s algorithm 
(which use is acknowledged here) for constructing multipole vectors from a standard spherical harmonic decomposition is described in \cite{copi2004} and the implementation of it is public available
\footnote{http://www.phys.cwru.edu/projects/mpvectors/}. 
Other methods exist but, as far as we know, their implementation is not public available
on a standard platform (see for example \cite{weeks,Katz} 
where the problem of finding $\ell$ vectors is translated into the problem of finding
the zeros of a polynomial of degree $2\ell$).

\subsection{Alignment estimators}
\label{alignmentestimators}

We focus on the alignment quadrupole-octupole.
Therefore we consider the following estimators widely used in literature
(e.g. \cite{weeks,abramo,Copi:2006,schwarz2004,copi2004}): 
\be
S23 &=& \sum_{i=1}^3|\vec q \cdot \vec o_i| /3  \, 
\label{defS23}
,\\
D23 &=& \sum_{i=1}^3|\hat q \cdot \hat o_i| /3
\, .
\label{defD23}
\ee
Here the symbol $\, \hat {} \, $ stands for a vector with norm equal to $1$.
The ``area vectors'' are defined as
\be
\vec q = \hat q_{2 1} \times \hat q_{2 2} \, , \\
\vec o_1 = \hat o_{3 2} \times \hat o_{3 3} \, , \\
\vec o_2 = \hat o_{3 3} \times \hat o_{3 1} \, ,\\
\vec o_3 = \hat o_{3 1} \times \hat o_{3 2} \, ,
\ee
where $\hat q_{2 j}$ represent the two normalized multipole vectors ($j=1,2$) associated to the quadrupole, 
$\hat o_{3 j}$ represent the three normalized multipole vectors ($j=1,2,3$) associated to the octupole.
 
Notice that all the estimators belong to the interval $[0,1]$ and contain absolute values 
in order to make them invariant under the reflection symmetry (see for example \cite{abramo,DSCgruppuso}).

\subsection{Description of the performed simulations and test for residuals}
\label{descriptions}

We have performed $3\times 10^5$ Gaussian random extractions of $\Lambda$CDM skies.
We have masked each extraction~\footnote{Hereafter we will use random extraction, random realization and random map as synonyms.}
setting the pixels
inside the mask to the fiducial value equal to $0$. 
This allows us to screen the information present in the Galactic region and, at the same time, 
to have an all-sky map (even if artificial) such that the multipole vectors
decomposition is still doable~\footnote{In this paper we will call masked 
maps, the maps that present the value 
$0$ for the pixels that fall into the mask, even if, strictly speaking, they are still all-sky.}.
We have largely exploited the extended temperature 
mask available at LAMBDA web site 
that has a sky coverage of $73\%$ 
and other two masks whose percentage of sky coverage are around $83\%$ and $93\%$
\footnote{We use "sky coverage" to mean "observed sky" (that has not to be confused with masked area).}.
For each realization (masked and not masked)
we have computed the estimators defined in Subsection \ref{alignmentestimators} 
passing from $a_{\ell m}$ to multipole vectors through the Copi et al.'s routine.
This allows us to compute the probability distribution function (henceforth pdf) of 
the considered estimators in the all-sky case and in the masked cases.
In all the cases (masked and unmasked) we have computed the value of
the estimators for four maps: the ILC map, the MCMC map, the V band map and the
foreground reduced V band
\footnote{A part the V band map that is taken into account as test case, all the other maps are only CMB maps obtained with different techniques. 
That's why we compare the same simulations with all these maps.}.
We do not use the masked pdf to directly extract scientific conclusions because this would not tell anything 
about the all sky map that is what we are interested in.
What we do use is the masking process itself. Here it is a description of the test we propose.
It consists of the following steps:
\begin{itemize}
\item consider the random maps whose all-sky estimator value belong
to the same bin of the pdf as the all-sky value of the estimator of the map under analysis;
\item mask these maps (both the set of random maps and the map under analysis that belong to the same bin) as described above;
\item transform to the multipole vectors representation and compute
the masked values of the estimator (both for the set of random maps and for the map under analysis that belong to the same bin);
\item build a pdf of these masked values obtained from the random extractions (what we call 
redistribution);
\item compare the masked value of the estimator for the map under analysis
with the redistribution.
\end{itemize}
In other words, we compute how each bin of interest of the all-sky pdf (for the considered estimator, D23 and S23 in this work) 
is redistributed when a mask is applied.
This allows us to analyze how consistent are the four considered maps with the 
random realizations whose D23 or S23 values belong to the same bin
of the all-sky distributions.
More explicitly, if the estimator of the masked map under analysis does not behave 
consistently with the other random masked maps then we conclude that some
residual was present in the all-sky map affecting the value of the all-sky estimator
and making it belonging to a ``wrong'' bin.
This consistency check is the test we propose in this paper to look for the effects 
of residuals, that are potentially present in the Galactic region
of the considered WMAP 5yr maps, on the alignment estimators.
Of course, in the case of the not cleaned V band map we are sure that 
a non-CMB signal is present at low Galactic latitudes. 

Note that the process adopted in our test
introduces a sort of {\it kernel} representing the effect of the mask
for each considered bin. 
The global pdf in the presence of a mask
can be seen as the convolution of the all-sky pdf with this kernel. 


The resolution that we have considered to perform this analysis
is represented by the HEALPix \footnote{http://healpix.jpl.nasa.gov}
 \cite{gorski} parameter $N_{side}$ that we have set~\footnote{For the reader who is not familiar with HEALPix convention, $N_{side}=16$
corresponds to maps of $3072$ pixels on the whole sky.} to $16$.

The impact of the noise in our test is addressed at the end of the next section.

\section{Results} 
\label{results}

In this Section we present and discuss the obtained results.
We have chosen the following convention, in order to display our results on the pdf:
red color for the all-sky pdf, cobalt blue color for the masked pdf,
green color for the difference between masked and unmasked pdf.
The green plots give the effect that has been
 introduced by the considered 
treatment of the mask.
More precisely they show the difference between masked 
and unmasked distribution (masked minus unmasked).
Moreover we use the blue color for the pdf of the redistribution for the large mask,
the light blue color for the pdf of the redistribution for the medium sized mask and
the white color for the pdf of the redistribution for the small mask.
Vertical lines show the values of the estimators derived from the considered maps:
black vertical line is for ILC map, blue vertical line for the MCMC map, green vertical line for the foreground reduced V band, yellow line for V band map.
The V band map has been taken into account just for comparison with the foreground reduced V band
and for method validation.

\subsection{Quadrupole-Octupole alignment estimators} 
\label{QO}

\begin{figure}
\includegraphics[width=90mm]{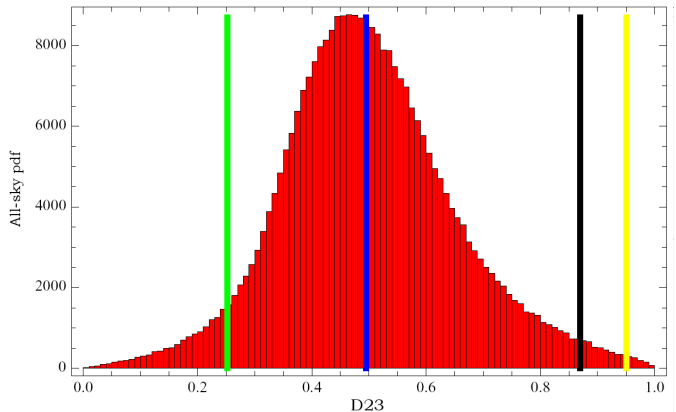}
\caption{All-sky pdf for D23. Vertical lines represent the values for
considered maps (of course in the all sky case). 
More precisely, 
black vertical line is for ILC map, blue vertical line for the MCMC map, green vertical line for the foreground reduced V band, yellow line for V band map.
The panel presents the counts (y-axis) versus the statistic (x-axis).
See also the text.}
\label{fullskyD23}
\end{figure}

\begin{figure*}

\includegraphics[width=16cm]{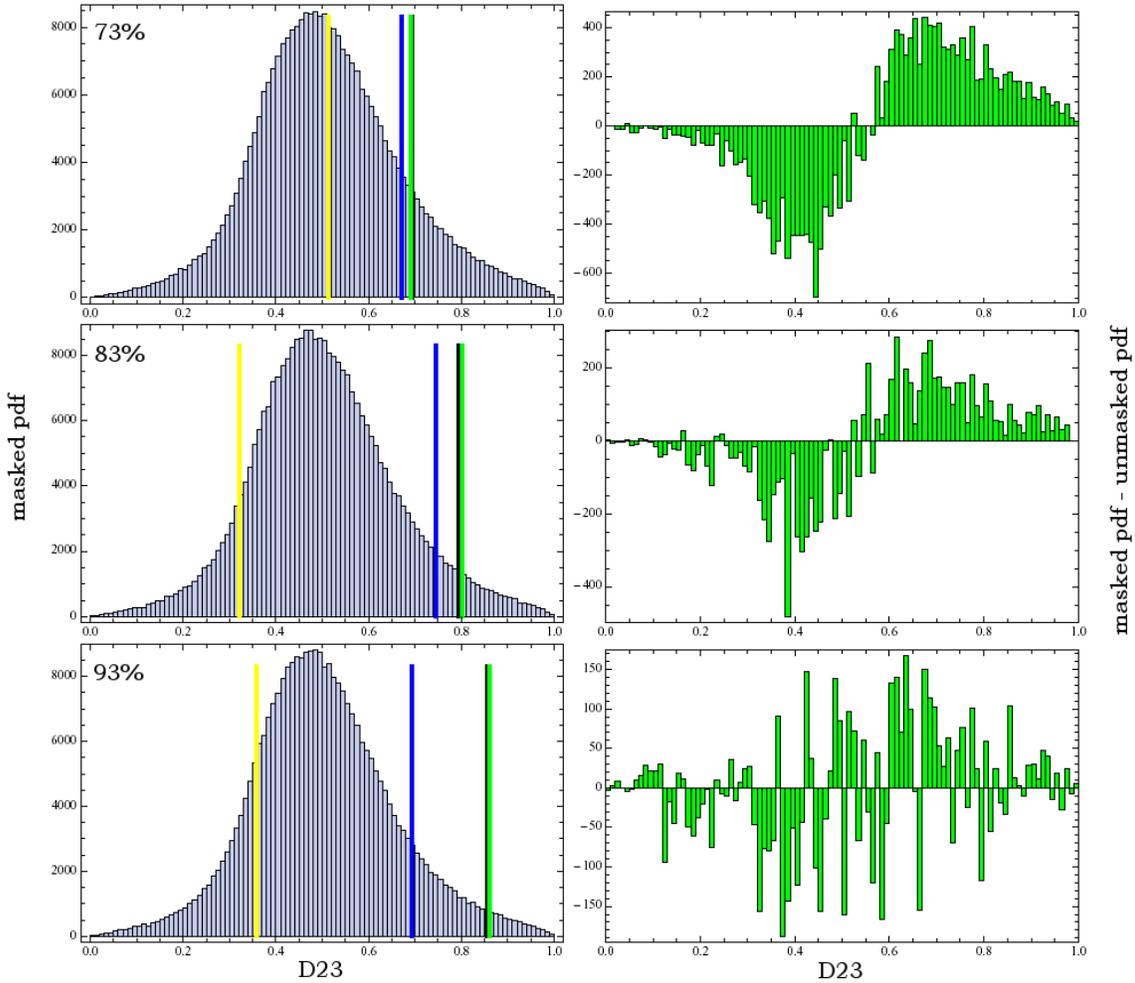}

%
%





\caption{Left column of panels: pdf of D23 for the masked case.
Black vertical line is for ILC map, blue vertical line for the MCMC map, green vertical line for the foreground reduced V band, yellow line for V band map.
Right column of panels: difference of the pdfs between masked and unmasked distributions. 
Percentage of sky coverage from top to bottom row of panels: $73\%$, $83\%$ and $93\%$.
All the panels present the counts (y-axis) versus the statistic (x-axis). See also the text.} 
\label{D23}
\end{figure*}

\begin{figure*}

\includegraphics[width=14.8cm]{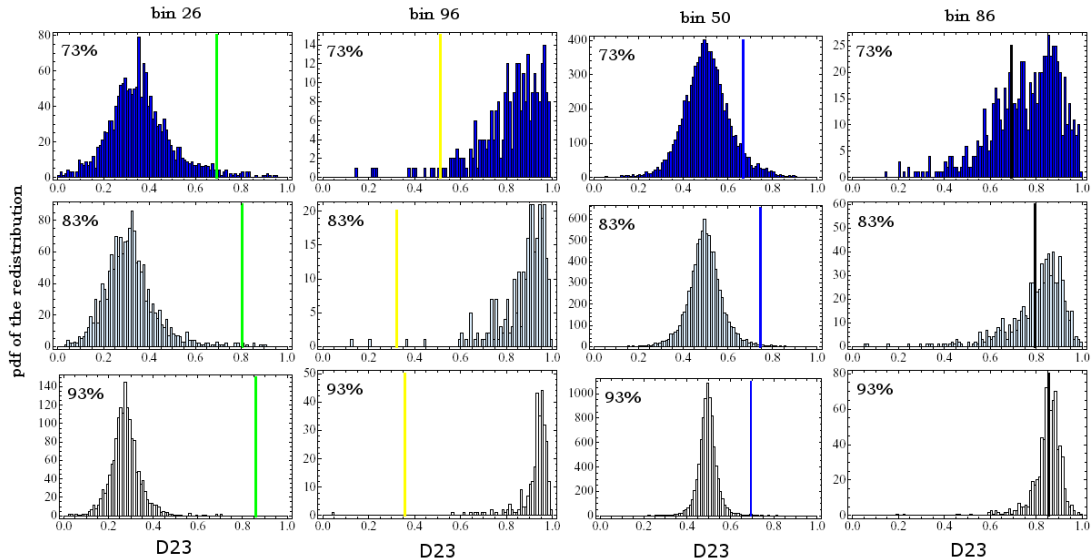}








\caption{Pdf of the redistribution of some specific bin of D23. In the first column we plot the redistribution due the masking process of the bin 26 for the foreground reduced V band map
(green vertical line). 
In the second column we plot the redistribution due the masking process of the bin 96 for the V band map (yellow vertical line).
In the third column we plot the redistribution due the masking process of the bin 50 for the MCMC map (blue vertical line).
In the fourth column we plot the redistribution due the masking process of the bin 86 for the ILC map (black line).
The first row is for the largest mask ($73\%$ of sky coverage), the second row for the middle size mask ($83\%$ of sky coverage) and the third row is for the smallest considered mask ($93\%$ of sky coverage).
All the panels present the counts (y-axis) versus the statistic (x-axis). See also the text.} 

\label{fluxD23}
\end{figure*}

\begin{figure*}

\includegraphics[width=14.4cm]{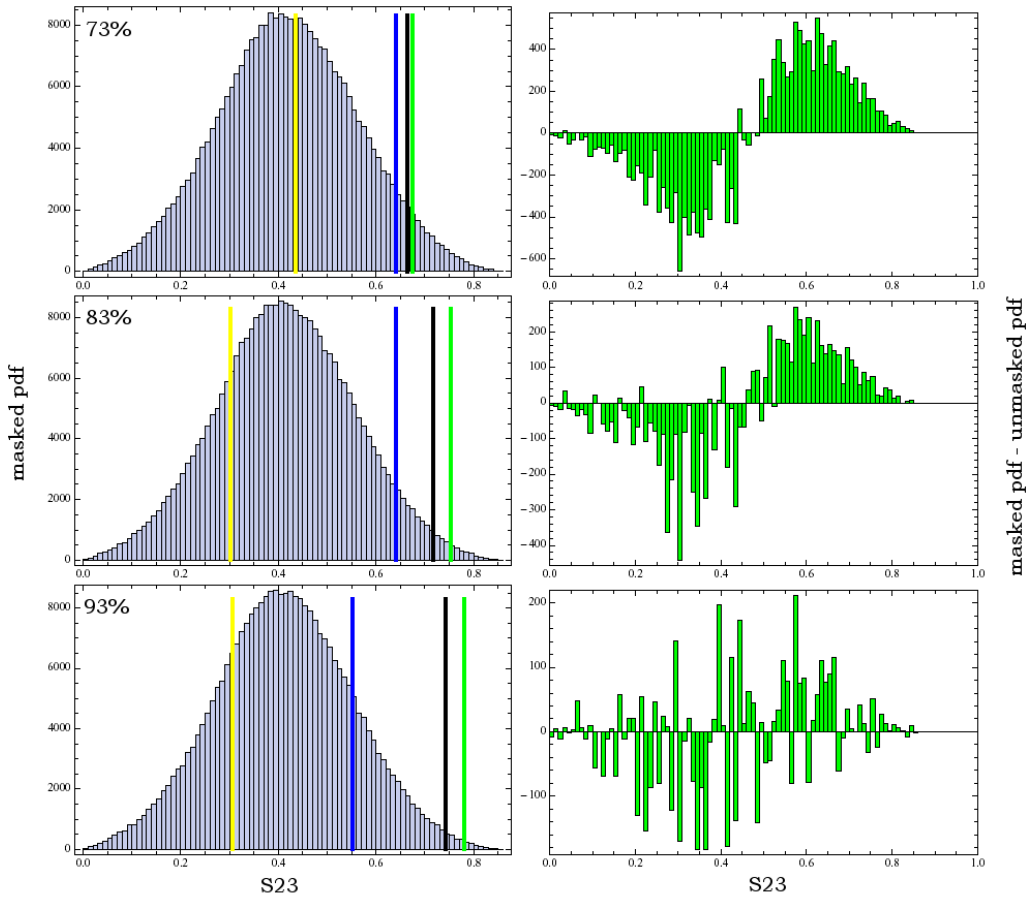}

\caption{The same as Fig. 3
but for S23. Left column of panels: pdf of S23 for the masked case.
Black vertical line is for ILC map, blue vertical line for the MCMC map, green vertical line for the foreground reduced V band, yellow line for V band map.
Right column of panels: difference of pdfs between masked and unmasked distributions. 
Percentage of sky coverage from top to bottom row of panels: $73\%$, $83\%$ and $93\%$.
All the panels present the counts (y-axis) versus the statistic (x-axis). See also the text.} 

\label{S23}

\end{figure*}

In Fig.~\ref{fullskyD23} we show the all sky pdf for D23 with the corresponding
values for the considered maps. 
In Fig.~\ref{D23} we show the same estimator D23 for the three considered masks
(left column of panels).
We note that for all the considered maps, but not for the V band map 
(yellow line) where 
foreground subtraction is not applied, the D23 estimator~\footnote{The same is true also for S23, see the following and Fig.~\ref{S23}.}
assumes similar values 
when intermediate or large masks are applied,
as the black, green, and blue lines tend to overlap 
(or at least to be closer) when the size of the Galactic mask increases.
In the right column of panels of Fig.~\ref{D23} there is the difference of the pdf's
between masked and all-sky case, 
that turns out to be stronger
for larger masks. 
At the chosen binning of $0.01$, such an effect appears weaker
for the smaller mask
\footnote{Increasing the number of realizations (for example to $10^6$) or the size of the bin (for example to $0.05$) would show
a similar shape as the other masked cases. 
We do not report here these plots but we dedicate a short appendix to this aspect.}.
It is interesting to notice that the area to the right of the black line (i.e. ILC map)
below the pdf is larger for larger masks.
In particular, the probability to obtain a value smaller than the ILC-WMAP value
is $98.36\%$ for the all-sky case, $97.96\%$ for smaller mask, $88.62\%$ for the medium mask,
$87.37\%$ for the extended mask. 
This would seem to indicate that excluding/screening the information coming from the Galactic region 
(as described in Section \ref{descriptions}) makes the estimator no longer anomalous
(for the map under analysis).
Unfortunately this does not indicate \emph{uniquely} that the anomaly that is present in the all-sky ILC-WMAP map is due to residuals that are present in the Galactic region. 
The fact that the masked ILC-WMAP map is no longer anomalous
might also be an effect of the masking process itself
(i.e. of the prior of setting to $0$ the masked pixels)
\footnote{Therefore, as aforementioned in footnote 4 and in Subsection \ref{descriptions}, we cannot extract scientific information about the all sky map from the masked analysis.}.
To discriminate between the two possibilities we compute the redistribution of the $87^{th}$ bin (that is the all-sky bin where the D23 value for the ILC-WMAP map falls) caused by the presence of zeros of the masking process. 
We analyze then if the masked ILC-WMAP D23 value is consistent with this redistribution, i.e. with a masked Gaussian random realization that exhibits a value for D23 that falls in same bin as the all-sky ILC-WMAP (this is the consistency check we propose).
%
%
Of course, the same has been done for the other maps and for all the three considered masks. 
Fig.~\ref{fluxD23} shows the ``masking flux''  of the bin 87 for the ILC-WMAP, of the bin 50 for the MCMC map, of the bin 96 for V band map and of the bin 26 for the foreground reduced V band map. 
Observing the fourth column of Fig.~\ref{fluxD23} we can see that the ILC-WMAP map is always consistent with a Gaussian random realization that shows the same all-sky D23 value.
For what concerns the other maps, Fig.~\ref{fluxD23} shows that they are not consistent with a Gaussian random realization because first, second and third column of Fig.~\ref{fluxD23}  present the D23 value for the other three maps (vertical lines) far from the peak of the distribution. 
We interpret this as evidence for the presence of residuals in the Galactic region in the all starting all-sky maps (except the aforementioned ILC-WMAP map).


\begin{figure}
\includegraphics[width=9.0cm]{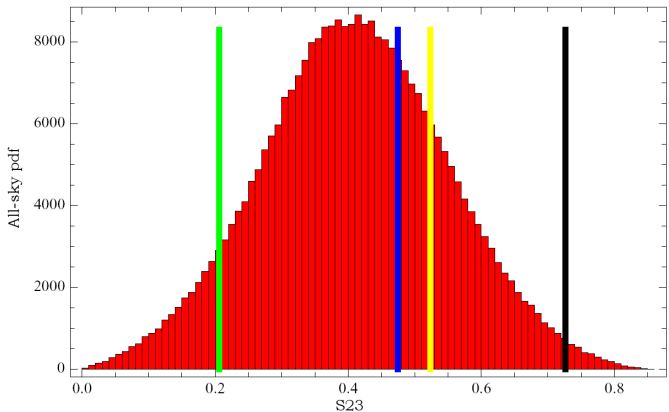}
\caption{The same as Fig.~2
but for S23.
Vertical lines represent the values for considered maps (of course 
in the all sky case).  More precisely, 
black vertical line is for ILC map, blue vertical line for the MCMC map, green vertical line for the foreground reduced V band, yellow line for V band map.
The panel presents the counts (y-axis) versus the statistic (x-axis). See also the text. }
\label{fullskyS23}
\end{figure}

\begin{figure*}

\includegraphics[width=14.8cm]{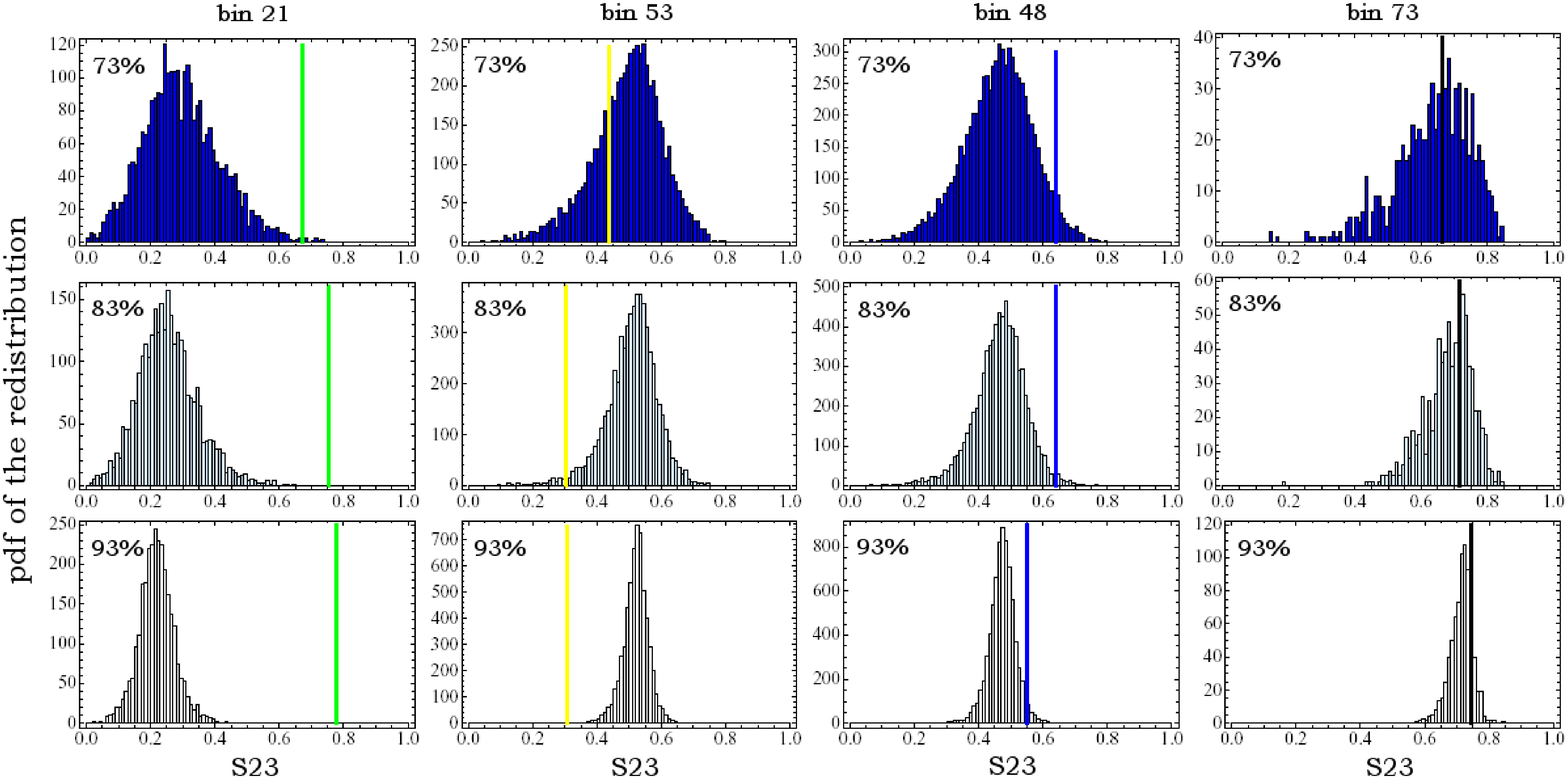}








\caption{Pdf of the redistribution of some specific bin of S23. In the first column we plot the redistribution due the masking process of the bin 21 for the foreground reduced V band map (green vertical line). 
In the same second column we plot the redistribution due the masking process of the bin 53 for the V band map (yellow vertical line).
In the third column we plot the redistribution due the masking process of the bin 48 for the MCMC map (blue vertical line).
In the fourth column we plot the redistribution due the masking process of the bin 73 for the ILC map (black vertical line).
The first row is for the largest mask ($73\%$ of the sky coverage), the second row for the middle size mask ($83\%$ of the sky coverage) and the third row is for the smallest considered mask ($93\%$ of the sky coverage).
All the panels present the counts (y-axis) versus the statistic (x-axis). See also the text.} 

\label{fluxS23}
\end{figure*}

In Fig.~\ref{fullskyS23} we show the all sky pdf for S23 with the corresponding
values for the considered maps. 
In Fig.~\ref{S23} we show the same estimator S23 for the three considered masks.
In particular, the probability to obtain a value smaller than the ILC-WMAP value
is $98.90\%$ for the all-sky case, $99.24\%$ for smaller mask, $98.50\%$ for the medium mask,
$95.44\%$ for the extended mask.
As for the D23 estimator, this does not {\it necessarily} mean 
presence of residuals coming from the Galactic region.
In Fig.~\ref{fluxS23} we show the redistribution for the bins where the S23 values of the considered maps stand.
In analogy with what has been obtained for the redistribution of the D23 bins, Fig.~\ref{fluxS23}
shows that the ILC-WMAP map is the only map that is consistent with the corresponding pdf of the
redistribution
\footnote{In fact the V band for the larger mask might be considered consistent, but the inconsistency become evident as soon as it is considered a less sized mask.
See also section \ref{conclusions}.}.
Note that the pdfs of redistributions of Figs.~\ref{fluxD23} and \ref{fluxS23} are more peaked as larger is the sky coverage. 

\begin{figure*}

\includegraphics[width=14.4cm]{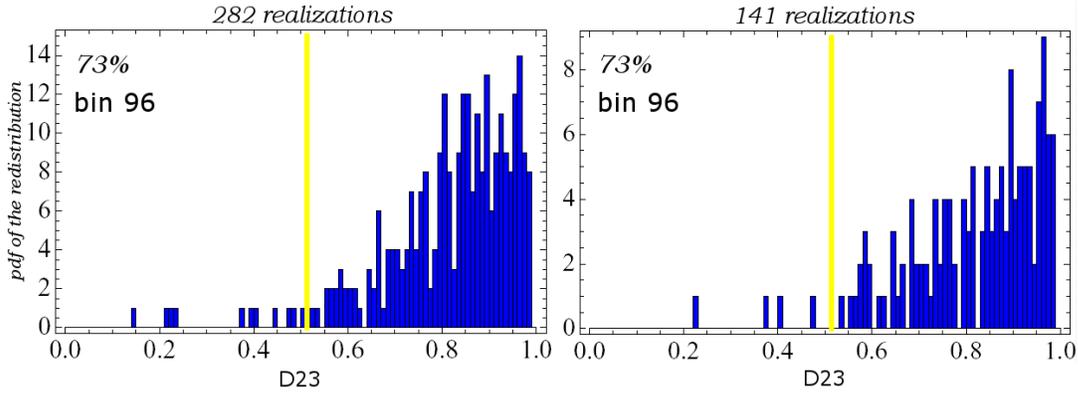}

\caption{Convergency of our results. Redistribution of D23 estimator, for bin 96 for the large mask case ($73 \%$ of sky coverage).
Left panel is exactly the same as in Fig.~4, while right panel uses half of the realizations belonging to the same bin.
The yellow vertical line represents the value of the estimator for the V band map. See also the text.} 

\label{population}

\end{figure*}

The redistributions shown in Figs.~\ref{fluxD23} and \ref{fluxS23} have well defined shapes.
The case of V band map and D23 estimator is the most anomalous (see Fig.~\ref{fullskyD23}), 
thus the corresponding bin is less populated ($282$ points). In spite of this, its redistributions are pretty stable.
In order to show this, we take the case of the large mask and compute in Fig.~\ref{population} the redistribution
for all the available points (left panel, as in Fig.~\ref{fluxD23}) and for half of the available points (right panel).
Since the shape appears to be stable for this case, the most critical one, we conclude that the shown results are robust.
In general, when the number of realization are few, an increase of the number of realizations and/or of the bin size of the histogram 
of the redistribution might be needed. 






Regarding the comparison between our estimators for the V band map
and the foreground reduced V band map, we note that the analysis of
the pdf of the redistribution in both cases 
always indicates the presence of residuals in the Galactic region,
except perhaps for the case of the S23 estimator 
applied to the V band map as suggested by the 
largest mask case~\footnote{This exception is not so 
surprising since the nature of our test
(note also that the CMB temperature anisotropy map in the V band is clearly not so 
affected by diffuse Galactic foregrounds when a large fraction of the 
sky at low Galactic latitudes is excluded).} (see Fig.~\ref{fluxS23}). 
Note that the power of our method depends also on the chosen estimator
and on the size of the mask. For this reason, given the estimator,   
we recommend to apply our test exploiting various mask sizes. 
Regarding the power of the method, in 
the considered cases, we found that the width of the pdf of 
redistribution increases (although not too strongly) with the size of
the mask. Note that for a map where we are aware of the presence of 
foregrounds, as the V band map, the test successfully detects 
spurious contamination.

\begin{figure}

\includegraphics[width=15.0cm]{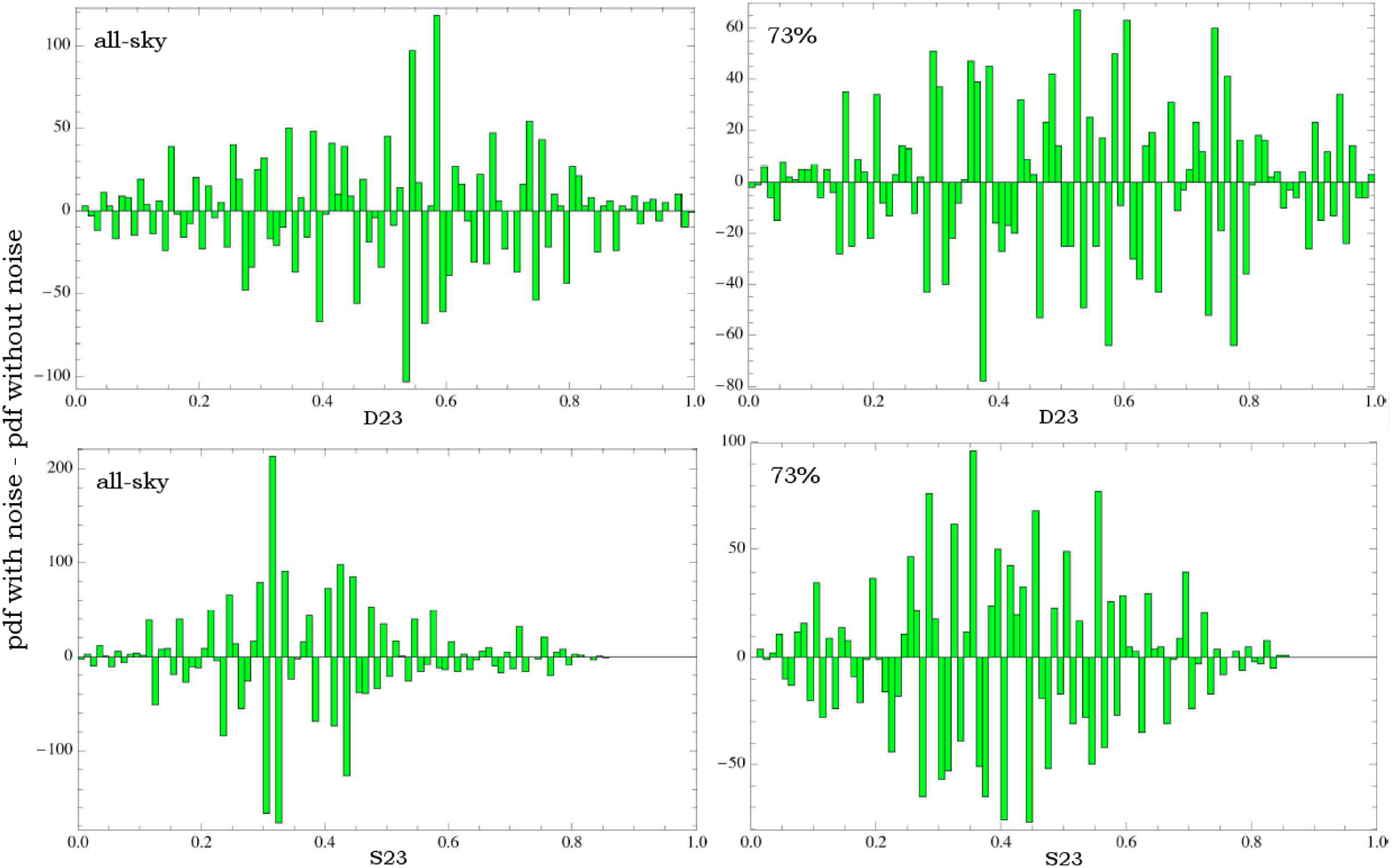}

\caption{Impact of the noise. Upper panels: difference of the pdf's for the D23 estimator.
Lower panels: difference of the pdf's for the S23 estimator.
Left panels: difference of the pdf in the case with noise minus without noise in the all-sky case.
Right panels: difference of the pdf in the case with noise minus without noise in the masked case
(extended mask, i.e. $73\%$ of sky coverage).
All the panels present the counts (y-axis) versus the statistic (x-axis). See also the text.} 
\label{noise}
\end{figure}

We have also studied the impact of the noise in the pdf of S23 and D23 in the masked ($73\%$ of sky coverage) and all-sky case. 
We expect only a (very) weak impact of the noise for our analysis as intuitively suggested by the fact that at $N_{side}=16$
the root mean square (RMS) of the signal (i.e. CMB) is $\simeq 50 \, \mu K$ and the RMS of the noise is
$ \simeq 2 \, \mu K$ (for example considering the V band map of WMAP).
In Fig.~\ref{noise} we show the effect of the noise on the pdf confirming the weak impact.
These plots have been obtained with $300000$ Gaussian random realizations. 
The CMB signal is extracted according to a $\Lambda$CDM model and the Gaussian random noise has been generated exploiting the non-uniform sensitivity of the V band map.

%

%



\section{Conclusions} 
\label{conclusions}

We have studied how the alignment between quadrupole and octupole is polluted on the Galactic region.


We found that among the considered maps the WMAP-ILC is the most consistent with Gaussian random realizations.
From the point of view of this test, we can say that WMAP-ILC is the most clean map (among the considered ones).
This result has been found studying the redistribution (due to the masking process) of each bin of the probability distribution functions of the alignment estimators.
Of course we cannot exclude that other tests might detect some residual contamination.
The used technique can indeed only exclude that the considered map is clean in some region but it cannot state that the map is clean in that region since a possible residual could be
in principle compatible with random realizations. 

Since the WMAP-ILC passes this test for three Galactic masks, the unlikely alignment between Quadrupole and Octupole (probed by the two considered estimators) that is present in this map is confirmed.

Although the results presented here formally apply to the considered products, our study points out on the relevance of analyzing low multipole alignments 
with the possibility of testing sky regions potentially affected by systematics (of instrumental or astrophysics origin) and show how it is feasible through Monte Carlo simulations. By construction, our method works for any possible mask adopted in the analysis of data from current and forthcoming CMB anisotropy experiments, such as those expected by the {\it Planck} satellite.

\acknowledgments
We acknowledge the use of the Legacy Archive for Microwave Background Data Analysis (LAMBDA). Support for LAMBDA is provided by the NASA Office of Space Science.
Some of the results in this paper have been derived using the HEALPix \cite{gorski} package.
We acknowledge the use of the public code for the multipole vectors decomposition
(quoted in footnote number $6$) \cite{copi2004}.
This work has been done in the framework of the {\it Planck} LFI activities.
We acknowledge the support by the ASI contracts ``{\it Planck} LFI Activity of Phase E2''
and I/016/07/0 ``COFIS''.

A particular thank to F.~Finelli for many useful discussions and our collaboration in this field.
We  acknowledge stimulating and fruitful conversations with P.~Bielewicz, K.~Gorski, E.~Martinez-Gonzalez, and P.~Vielva.
We thank P.~Procopio and J.~Zuccarelli for their kind graphic help.
We acknowledge the anonymous Referee for constructive comments.

\appendix

\section{Comparison of simulations with more realizations}
\label{Comparison of simulations with more realizations}
\begin{figure}

\includegraphics[width=15.0cm]{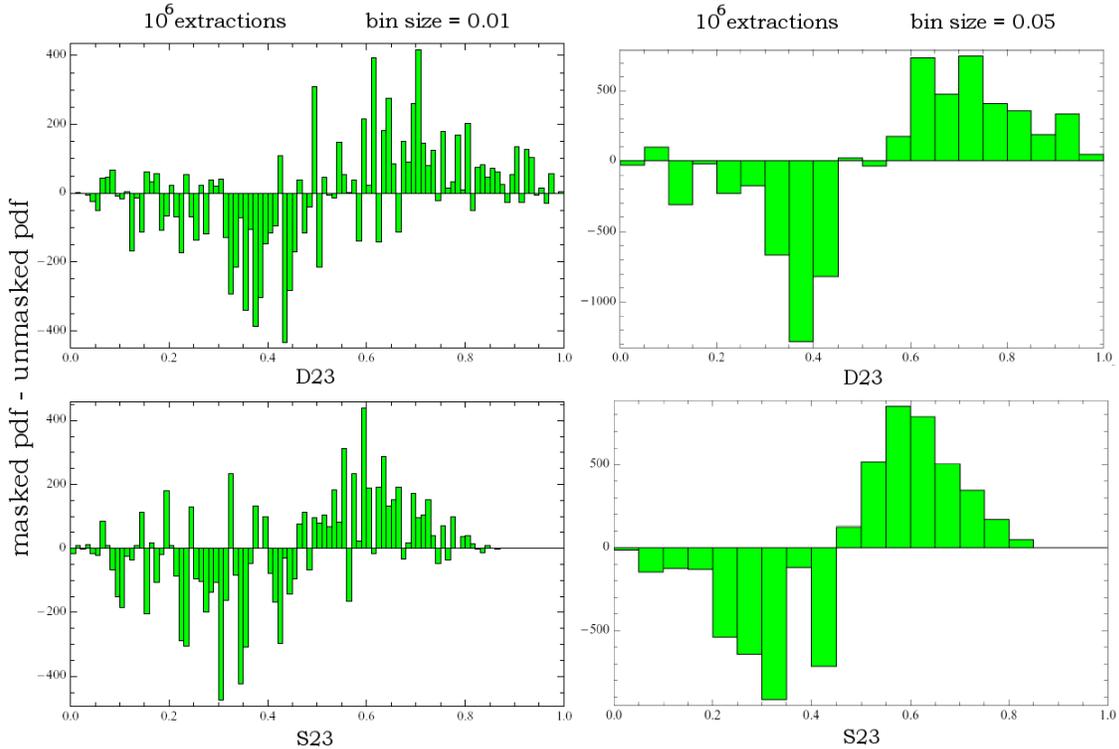}

\caption{Impact of the number of realizations on the considered estimators for the smaller mask case ($93\%$ of sky coverage). 
All the panels are referring to the smaller mask case ($93\%$ of sky coverage). Upper panels: difference of the pdfs 
between masked and full sky of D23 for the masked case.
Lower panels: difference of the pdfs between masked and full sky of S23 for the masked case.
Left panels: $1000000$ extractions with binning of $0.01$. Right panel: $1000000$ extractions and 
binning of $0.05$. 
See lower-right panels of Figs 3
and 6
for comparison with the corresponding pdf built with $300000$ extractions. 
All the panels present the counts (y-axis) versus the statistic (x-axis). See also the text.} 
\label{numberextraction}
\end{figure}

In Fig.~\ref{numberextraction} we focus on the smallest mask case. This case for the difference of pdfs 
seems not to have a clear shape (see lower-right panels of Fig.~\ref{D23} and \ref{S23}). Therefore we increased the number of extractions in order to improve the statistics. 
In Fig.~\ref{numberextraction} we show the results for differences of the pdf's between masked and unmasked distribution of D23 and S23 with $1000000$ extractions with binning of $0.01$ and $0.05$. 
This demonstrates that even the smaller mask case exhibits qualitatively the same behavior of the other cases (even if weaker quantitatively).
Anyway, we suggest to use larger number of realizations when small masks are exploited.

\begin{thebibliography}{999}

\bibitem{dunkley} 
Dunkley J. {\it et al.}  [WMAP Collaboration], 2008,
arXiv:0803.0586 [astro-ph].
\bibitem{komatsu} 
Komatsu E. {\it et al.}  [WMAP Collaboration], 2008,
arXiv:0803.0547 [astro-ph].
\bibitem{Copi2008}
Copi C.J., Huterer D., Schwarz D.J. and Starkman G.D., 2008,
arXiv:0808.3767 [astro-ph].
\bibitem{Copi2006}
Copi C.J., Huterer D., Schwarz D.J. and Starkman G.D., 2007,
  Phys.\ Rev.\  D {\bf 75}, 023507
\bibitem{Smoot1992}
Smoot G.F. {\it et al.}, 1992,
Astrophys.\ J.\  {\bf 396}, L1
\bibitem{Hinshaw1996}
 Hinshaw G., Banday A.J., Bennett C.L., Gorski K.M., Kogut A., Smoot G.F. and Wright E.L., 1996,
arXiv:astro-ph/9601058.
\bibitem{copi2004}
Copi C.J., Huterer D. and Starkman G.D., 2004,
Phys.\ Rev.\  D {\bf 70}, 043515
\bibitem{weeks}
Weeks J.R., 2004,
arXiv:astro-ph/0412231.
\bibitem{tegmark2003}
Tegmark M., de Oliveira-Costa A. and Hamilton A., 2003,
Phys.\ Rev.\  D {\bf 68}, 123523
\bibitem{schwarz2004}
Schwarz D.J., Starkman G.D., Huterer D. and Copi C.J., 2004,
Phys.\ Rev.\ Lett.\  {\bf 93}, 221301
\bibitem{land2005}
Land K. and Magueijo J., 2005,
Phys.\ Rev.\ Lett.\  {\bf 95}, 071301
\bibitem{vale2005}
Vale C., 2005,
arXiv:astro-ph/0509039.
\bibitem{abramo}
Abramo L. R., Bernui A., Ferreira I.S., Villela T. and C. Wuensche C.A., 2006,
  Phys.\ Rev.\  D {\bf 74}, 063506
\bibitem{wiaux}
 Wiaux Y., Vielva P., Martinez-Gonzalez E. and Vandergheynst P.,
  Phys.\ Rev.\ Lett.\  {\bf 96}, 151303 (2006)
  [arXiv:astro-ph/0603367].
\bibitem{vielva2007}
Vielva, P., Wiaux, Y., Martõnez-Gonzalez, E., Vandergheynst, P.,
 Mon.\ Not.\ Roy.\ Astron.\ Soc.\  {\bf 381}, 932 (2007)
\bibitem{Eriksen}
Eriksen H.K., Hansen F.K., Banday A.J., Gorski K.M. and Lilje P.B., 2004,
Astrophys.\ J.\  {\bf 605}, 14
[Erratum-ibid.\  {\bf 609}, 1198]
\bibitem{hansen}
Hansen F.~K., Banday A.~J. and Gorski K.~M.,
  Mon.\ Not.\ Roy.\ Astron.\ Soc.\  {\bf 354}, 641 (2004)
  [arXiv:astro-ph/0404206].
 \bibitem{vielva}
  Vielva P., Martinez-Gonzalez E., Barreiro R.~B., Sanz J.~L. and Cayon L.,
  Astrophys.\ J.\  {\bf 609}, 22 (2004)
  [arXiv:astro-ph/0310273].
\bibitem{cruz}
  Cruz M., Martinez-Gonzalez E., Vielva P. and Cayon L.,
  Mon.\ Not.\ Roy.\ Astron.\ Soc.\  {\bf 356}, 29 (2005)
  [arXiv:astro-ph/0405341].
\bibitem{abramo2}
Abramo L.R., Jr L.S. and Wuensche C.A., 2006,
Phys.\ Rev.\  D {\bf 74}, 083515 
\bibitem{Naselsky:2006mt}
Naselsky P.D. and Verkhodanov O.V., 2008,
Int.\ J.\ Mod.\ Phys.\  D {\bf 17}, 179
\bibitem{Inoue2006}
Inoue K.T. and Silk J., 2006,
Astrophys.\ J.\  {\bf 648}, 23
\bibitem{Inoue2007}
Inoue K.T. and Silk J., 2007,
Astrophys.\ J.\  {\bf 664}, 650
\bibitem{cooray2005}
Cooray A. and Seto N., 2005,
JCAP {\bf 0512}, 004
\bibitem{DSCburigana}
Burigana C., Gruppuso A. and Finelli F., 2006,
  Mon.\ Not.\ Roy.\ Astron.\ Soc.\  {\bf 371}, 1570
\bibitem{DSCgruppuso}
Gruppuso A., Burigana C. and Finelli F., 2007,
Mon.\ Not.\ Roy.\ Astron.\ Soc.\  {\bf 376}, 907
\bibitem{Coles:2003dw}
Coles P, Dineen P., Earl J. and Wright D., 2004,
Mon.\ Not.\ Roy.\ Astron.\ Soc.\  {\bf 350}, 983
\bibitem{Park:2006dv}
Park C.G., Park C. and Gott J.R.I., 2007,
Astrophys.\ J.\  {\bf 660}, 959
\bibitem{Naselsky:2007gt}
Naselsky P.D., Verkhodanov O.V. and Nielsen M.T.B., 2008,
Astrophys.\ Bull.\  {\bf 63}, 216
\bibitem{Chiang:2007rp}
Chiang L.Y., Naselsky P.D. and Coles P., 2008,
Mod.\ Phys.\ Lett.\  A {\bf 23}, 1489
\bibitem{Luminet:2003dx}
Luminet J.P., Weeks J., Riazuelo A., Lehoucq R. and Uzan J.P., 2003,
Nature {\bf 425}, 593
\bibitem{Weeks:2006rr}
Weeks J.R. and Gundermann J., 2007,
Class.\ Quant.\ Grav.\  {\bf 24}, 1863
\bibitem{Contaldi:2003zv}
Contaldi C.R., Peloso M., Kofman L. and Linde A., 2003,
JCAP {\bf 0307} 002
\bibitem{Moroi:2003pq}
Moroi T. and Takahashi T., 2004,
Phys.\ Rev.\ Lett.\  {\bf 92} 091301
\bibitem{Campanelli2006}
Campanelli L., Cea P. and L.~Tedesco L., 2006,
Phys.\ Rev.\ Lett.\  {\bf 97}, 131302
[Erratum-ibid.\  {\bf 97}, 209903]
\bibitem{Gruppuso2007}
Gruppuso A., 2007,
Phys.\ Rev.\  D {\bf 76}, 083010
\bibitem{Finelli:2005zc}
Finelli F. and Gruppuso A., 2005,
arXiv:hep-th/0501089.
\bibitem{Barrow1997}
Barrow J.D., Ferreira P.G. and Silk J., 1997,
Phys.\ Rev.\ Lett.\  {\bf 78}, 3610
\bibitem{Gosh2007}
Ghosh T., Hajian A. and Souradeep T., 2007
Phys.\ Rev.\  D {\bf 75}, 083007
\bibitem{Jaffe2006}
Jaffe T.R., Banday A.J., Eriksen H.K., Gorski K.M. and Hansen F.K., 2006,
Astron.\ Astrophys.\  {\bf 460}, 393
\bibitem{de OliveiraCosta:2003pu}
de Oliveira-Costa A., Tegmark M., Zaldarriaga M. and Hamilton A., 2004,
Phys.\ Rev.\  D {\bf 69}, 063516
\bibitem{Bunn}
Bunn E.F. and Bourdon A., 2008,
arXiv:0808.0341 [astro-ph].
\bibitem{bielewicz_2}
Bielewicz P., Eriksen H.K., Banday A.J., Gorski K.M. and Lilje P.B., 2005,
Astrophys.\ J.\  {\bf 635}, 750
\bibitem{bielewicz_1}
Bielewicz P., Gorski K.M. and Banday A.J., 2004,
Mon.\ Not.\ Roy.\ Astron.\ Soc.\  {\bf 355}, 1283
\bibitem{deOliveiraCosta:2006}
de Oliveira-Costa A.and Tegmark M., 2006,
Phys.\ Rev.\  D {\bf 74}, 023005
\bibitem{maxwell}
Maxwell J.C., 1873, ``A Treatise on Electricity and Magnetism''
reprinted by Dover Publication, 1954
\bibitem{sakurai}
Sakurai J.J., 1985, ``Modern Quantum Mechanics''. Revised
Edition. Addison-Wesley Publishing Company, Inc.
\bibitem{landmagueijo}
Land K. and Magueijo J., 2005,
Mon.\ Not.\ Roy.\ Astron.\ Soc.\  {\bf 362}, 838
\bibitem{dennis}
Dennis M.R., 2004, 2005,
J. Phys. A 37 9487-9500, 
J.Phys. A38 1653-1658
\bibitem{Katz}
Katz G. and Weeks J., 2004,
Phys.\ Rev.\  D {\bf 70}, 063527
\bibitem{Copi:2006}
  Copi C.J., Huterer D., Schwarz D.J. and Starkman G.D.,
  Mon.\ Not.\ Roy.\ Astron.\ Soc.\  {\bf 367}, 79 (2006)
\bibitem{gorski}
Gorski K.M., Hivon E., Banday A.J., Wandelt B.D., Hansen F.K., Reinecke M. and Bartelmann M., 2005,
HEALPix: A Framework for High-resolution Discretization and Fast Analysis of Data Distributed on the Sphere, Ap.J., 622, 759-771










\end{thebibliography}
\end{document}